\documentclass[preprint]{aastex}

\slugcomment{Submitted to the Astrophysical Journal}
\received{2/1/2001}
\revised{4/20/2001}
\accepted{4/24/2001}

\makeatletter
\def\ale{\mathrel{\mathpalette\gl@align<}}
\def\age{\mathrel{\mathpalette\gl@align>}}
\def\gl@align#1#2{\lower.6ex\vbox{\baselineskip\z@skip\lineskip\z@
\ialign{$\m@th#1\hfil##\hfil$\crcr#2\crcr\sim\crcr}}}
\makeatletter
\newcommand{\iras}{{\sl IRAS\/ }}

\newcommand{\um}{\mu m}
\newcommand{\teff}{T_{\rm eff}}
\newcommand{\mdot}{{\dot M}}
\newcommand{\msun}{M_{\odot}}
\newcommand{\lsun}{L_{\odot}}

\newcommand{\lstar}{L_{*}}
\newcommand{\thincl}{\theta_{\rm incl}}

\shorttitle{Mid-IR Emission Structure of IRAS 22272+5435} 
\shortauthors{Ueta et al.}

\begin{document}
 
\title{Sub-arcsecond Mid-IR Structure of the Dust Shell around 
IRAS 22272+5435\footnote{The observational data presented here 
were obtained at the MMT Observatory, a joint facility of the 
University of Arizona and the Smithsonian Institution, and
at the W. M. Keck Observatory, which was made possible
by the generous financial support of the W. M. Keck Foundation
and is operated as a scientific partnership among the California 
Institute of Technology, the University of California and the 
National Aeronautics and Space Administration.}}

\author{Toshiya Ueta\altaffilmark{1}, 
Margaret Meixner\altaffilmark{1}, 
Philip M. Hinz\altaffilmark{2}, 
William F. Hoffmann\altaffilmark{2},
Wolfgang Brandner\altaffilmark{3},
Aditya Dayal\altaffilmark{4},
Lynne K. Deutsch\altaffilmark{5},
Giovanni G. Fazio\altaffilmark{6}, and
Joseph L. Hora\altaffilmark{6}}

\altaffiltext{1}{Department of Astronomy, MC-221, 
University of Illinois at Urbana-Champaign, 
Urbana, IL  61801; 
ueta@astro.uiuc.edu, meixner@astro.uiuc.edu}

\altaffiltext{2}{Steward Observatory, 
University of Arizona, 
Tucson, AZ  85721;
phinz@as.arizona.edu, whoffmann@as.arizona.edu}

\altaffiltext{3}{Institute for Astronomy, 
University of Hawaii,
2680 Woodlawn Drive,
Honolulu, HI 96822;
brandner@ifa.hawaii.edu}

\altaffiltext{4}{KLA-Tencor Corp.,
160 Rio Robles,
San Jose, CA 95134; 
Aditya.Dayal@kla-tencor.com}

\altaffiltext{5}{Department of Astronomy/CAS 519, 
Boston University, 
725 Commonwealth Avenue, 
Boston, MA  02215;
deutschl@bu.edu}

\altaffiltext{6}{Harvard-Smithsonian Center for Astrophysics, MS 65,
60 Garden St., 
Cambridge, MA  02138;
jhora@cfa.harvard.edu, gfazio@cfa.harvard.edu}

\begin{abstract}
We report sub-arcsecond imaging of extended mid-infrared 
emission from a proto-planetary nebula (PPN), \iras 22272+5435,
performed at the MMT observatory with its newly upgraded 6.5 m 
aperture telescope and at the Keck observatory.
The mid-infrared emission structure is resolved into two
emission peaks separated by $0\arcsec.5 - 0\arcsec.6$ 
in the MMT 11.7 $\um$ image and in the Keck 7.9, 9.7, and 
12.5 $\um$ images, 
corroborating the predictions based on previous 
multi-wavelength morphological studies and radiative 
transfer calculations.
The resolved images show that the PPN dust shell
has a toroidal structure with the $0\arcsec.5$ inner 
radius.
In addition, an unresolved mid-IR excess appears at 
the nebula center.
Radiative transfer model calculations suggest that 
the highly equatorially-enhanced
($\rho_{\rm eq}/\rho_{\rm pole} = 9$) structure of 
the PPN shell was generated by  
an axisymmetric superwind with
${\dot M}_{\rm sw} = 4 \times 10^{-6} M_{\odot}$ yr$^{-1}$,
which was preceded by a spherical asymptotic giant 
branch (AGB) wind with
${\dot M}_{\rm AGB} = 8 \times 10^{-7} M_{\odot}$ yr$^{-1}$.
These model calculations also indicate that the 
superwind shell contains larger dust grains than
the AGB wind shell.
The unresolved mid-infrared excess may have been 
produced by a post-AGB mass loss at a rate of 
$2 - 6 \times 10^{-7} M_{\odot}$ yr$^{-1}$.
While the central star left the AGB about 380 years ago
after the termination of the superwind,
the star seems to have been experiencing an ambient
post-AGB mass loss with a sudden, increased mass 
ejection about 10 years ago.
\end{abstract}


\section{Introduction}

Proto-planetary nebulae (PPNs) are evolved stars of 
low-to-intermediate initial mass ($0.8 - 8 \msun$) 
that are in transition from the asymptotic giant branch 
(AGB) phase to the planetary nebula (PN) phase
(e.g., \citealt{kwok93,iben95}).
A PPN is a stellar system comprising of the central 
star of B - K spectral type surrounded by a detached 
circumstellar dust/gas shell.
Therefore, PPNs are often bright at both optical and 
infrared (IR) wavelengths, resulting in the 
``double-peaked'' spectral energy distribution (SED) 
- a characteristic signature of these post-AGB objects 
\citep{vhg89,kwok93}.
This particularly strong IR excess of a PPN is 
primarily due to thermal emission arising from
the circumstellar dust shell, which is created
by mass loss in the AGB phase.

The morphologies of the PPN dust shells are found to be
predominantly axisymmetric
(e.g., \citealt{ueta00} and reference therein),
and such axisymmetric PPN shells are
generally believed to interact with a fast post-AGB 
wind to form spectacularly aspherical PN shells
by the time ionization of the shell material
begins to take place
(the interacting stellar wind model; \citealt{kwok82}).
Thus, PPN dust shells retain the most pristine,
almost complete AGB mass loss histories 
that have not yet been altered by the interacting 
winds and/or energetic stellar radiation. 
We can, therefore, address the issues such as the cause 
of the morphological transformation in the circumstellar 
shells and the nature of the post-AGB wind
by closely observing the PPN shells, especially
at the innermost region of the PPN shells.

In this context, a variety of high-resolution 
observational techniques has been employed to 
observe the PPN dust shells.
For example, high-resolution optical imaging has been 
very effective at unveiling the remarkably axisymmetric 
structure of the PPN reflection nebulosities
(e.g., \citealt{sahai98,kwok98,su98,hrivnak99,ueta00}).
While such optical images of superb resolution uncovered
detailed geometrical information of the PPN shells, 
optical emission, i.e., dust-scattered star light, allows 
only an indirect probe of the dust mass distribution.
Therefore, mid-IR imaging, which directly probes the 
mass distribution through thermal emission arising from 
dust grains, has also been employed
(e.g., \citealt{skinner94,meixner97,meixner99}).
However, the diffraction-limited nature of mid-IR
observations permits imaging only at marginal 
($\sim 1\arcsec$) resolution with existing 3m class 
IR telescopes.

With large aperture telescopes coming on-line, 
it is now becoming possible to obtain sub-arcsecond
resolution mid-IR images (e.g., \citealt{jura99,jura01,morris00})
which would further improve our understanding of
the axisymmetric shaping in post-AGB objects.
In this paper, we report sub-arcsecond mid-IR 
imaging of a PPN, \iras 22272+5435, and 
the results of 2-D radiative transfer calculations 
based on the spatial information obtained from the
high-resolution mid-IR images.

\section{The Object: \iras 22272+5435}

\iras 22272+5435 ($=$ HD 235858 $=$ SAO 34504; hereafter 
\iras 22272) was identified as a PPN candidate
soon after its detection 
by the {\sl Infrared Astronomical Satellite} ({\sl IRAS})
because of the fairly bright nature both in the optical
(${\sl V\/}= 8.9$) and IR \citep{pottasch88}.
Subsequent spectroscopic observations classified the 
central star as G5 and revealed its carbon-rich (C$-$rich) 
nature through molecular carbon features 
(\citealt{hrivnak91, hrivnak95}).
The C$-$rich nature of the source was 
also shown by detection of various carbon-bearing 
molecular species \citep{lind88,omont93}
and the IR hydrocarbon features
(\citealt{buss90, geballe92, justtanont96}).
Moreover, 
\citet{zacs95} have recently revealed 
an extremely C$-$rich nature (C/O $\sim 12$) of the source
by high-resolution spectroscopic observations.

In addition to the IR hydrocarbon features at around 
$5 - 13$ $\um$, \iras 22272 is known to have strong 
IR features near 21 and 30 $\um$, whose carrier 
species have not been firmly identified.
The 21 $\um$ feature was originally discovered 
as a result of an investigation in the \iras low 
resolution spectrometer (LRS) database
(\citealt{kvh89}), and subsequent investigation
found that the 21 $\um$ feature seemed to be
a unique feature of C$-$rich PPNs \citep{kvh99}.
A recent {\sl ISO} spectroscopic survey of the 
21 $\um$ sources has determined the intrinsic 
shape of the feature that would suggest the 
solid-state nature of the carrier species 
(\citealt{vkh99}).
However, the origin of the feature remains elusive
despite a number of suggestions for its carrier
species (e.g., \citealt{begeman96,justtanont96,hill98,helden00}).
The 30 $\um$ feature was discovered by \citet{omont95}
during their spectroscopic surveys of C$-$rich PPNs 
to look for additional IR features which would help
to identify the 21 $\um$ feature.
The 30 $\um$ feature in \iras 22272 is enormous and
the integrated flux over the feature accounts for roughly
$20\%$ of the bolometric luminosity \citep{omont95}.
Because the 30 $\um$ feature is also seen in C$-$rich 
AGB stars and PNs, the 30 $\um$ does not seem to 
share the transient nature of the 21 $\um$ carrier
species.
The identification of the 30 $\um$ carrier species 
is also still incomplete:
suggested species include, for example, MgS 
\citep{goebel85} and a mixture of hydrocarbons 
\citep{duley00}.

Morphologies of \iras 22272 have been studied previously 
at various wavelengths.
\citet{meixner97} and \citet{dayal98} presented
mid-IR images at arcsecond resolution which showed 
clear elongation in the emission core.
These authors respectively suggested the presence of an 
inclined dust torus or a disk by means of radiative transfer 
calculations and the color temperature and optical 
depth maps derived from the source images.
High-resolution optical images obtained by
{\sl Hubble Space Telescope} ({\sl HST}) revealed a 
reflection nebulosity of very faint surface brightness 
with a clear view of the central star at the center of 
the nebula \citep{ueta00}.
The optical nebulosity was largely elongated perpendicular
to the core elongation seen at the mid-IR, suggesting
that the preferred directions of the dust-scattered 
star light are defined by the biconical openings of 
the suspected dust torus.
A recent near-IR imaging by \citet{gledhill00}
was able to separate polarized (i.e., dust-scattered) 
emission arising from the dust grains from the 
unpolarized direct stellar emission by means of 
imaging polarimetry.
Their {\sl J\/} band polarized flux image showed a
ring-like structure embedded in an elongated halo,
corroborating the findings of the mid-IR and optical 
imaging.

Based on the combined results of our recent mid-IR 
and {\sl HST} imaging surveys of PPNs 
\citep{meixner99,ueta00}, 
\iras 22272 is morphologically classified as a 
SOLE-toroidal PPN.
The SOLE-toroidal morphological type is considered 
to be caused by an optically {\sl thin} dust torus:
the dust-scattered star light is only marginally 
confined in the biconical openings thereby forming 
an elliptically elongated optical
nebulosity with a clear view of the central star
in the middle of the nebula,
while the mid-IR thermal emission from dust grains
shows some evidence for the central dust torus
through either an elongation of the core or 
two emission peaks when resolved 
(e.g., Figure 5 of \citealt{ueta00}).
Especially important is the resolved two-peak 
structure of the mid-IR emission core because 
these peaks are manifestations of the limb-brightened 
edges of an optically {\sl thin} dust torus, 
acting as the direct evidence for the 
toroidal dust distribution in the PPN shell.
However, because of the diffraction-limited mid-IR 
imaging and the intrinsically compact nature of 
the PPN dust shells, 
only 22 out of 72 PPNs have been resolved in the 
mid-IR so far.
Moreover, there have been only two out of 
11 SOLE-toroidal PPNs in which the central dust
torus is well-resolved to show two limb-brightened
peaks in the mid-IR emission core%
\footnote{These two cases are \iras 17436+5003
\citep{skinner94} and \iras 07134+1005
\citep{meixner97,jura01}.  
See also Figure 5 of \citet{ueta00}.}.

Now, \iras 22272 is known to have the largest 
elongated emission core 
(1\arcsec.6 diameter at 11.8 $\um$; \citealt{meixner97})
among the SOLE-toroidal PPNs in which the two-peak
core structure has not been resolved.
Using the arcsecond resolution images,
\citet{meixner97} have estimated the inner radius of the 
central dust torus to be about $0\arcsec.6$ by
2-D radiative transfer calculations.
Therefore, \iras 22272 is the prime target PPN
for the mid-IR imaging at sub-arcsecond resolution
with a large aperture telescope, and 
we conducted imaging of the source at 
the MMT and W. M. Keck Observatories using two
mid-IR cameras in order to resolve the yet unresolved
two limb-brightened edges in the emission core.
Recently, the presence of more than one emission peaks 
has been tentatively confirmed by an 8.8 $\um$ image 
taken at the Keck telescope which shows two peaks in the mid-IR
emission nebula \citep{morris00}.

\section{Sub-arcsecond Mid-IR Imaging and Results}

\subsection{Observations and Data Reduction}

We observed \iras 22272 at the MMT observatory on 2000 June 16 
during the engineering/commissioning period that followed its 
successful first-light on 2000 May 17 after the primary mirror 
conversion \citep{west97}.
The observations were made under clear sky 
with the University of Arizona/Smithsonian Astrophysical 
Observatory Mid-IR Array Camera 
(MIRAC3; \citealt{hoffmann98}) interfaced with the University of 
Arizona BracewelL Infrared Nulling Cryostat (BLINC; 
\citealt{hinz98}) mounted at the Cassegrain focus of the 
upgraded 6.5m primary mirror.
The MIRAC3 array is a Boeing HF-16 arsenic-doped silicon 
blocked-impurity-band hybrid array and has a $128 \times 128$ 
pixel format, which would give a $19\arcsec \times 19\arcsec$ 
field of view with the $0\arcsec.15$ pixel scale.
The object was observed with $10\%$ bandwidth 
($\Delta \lambda/\lambda = 0.1$) filters at 8.8, 9.8 and 11.7 
$\um$ without the BLINC nulling mode.

Because the f/9 secondary - the only available secondary for
the commissioning period - was not capable of chopping  
or tip-tilt, 
a nod-only beam-switching was used with an elevation nod throw 
of 9\arcsec, which yielded two on-chip sky-subtracted
images of the source in a single integration cycle.
The telescope was experiencing a drift in the azimuth direction 
with a rate of approximately 0\arcsec.2 min$^{-1}$
in spite of its excellent pointing accuracy.
Therefore, we were limited to the maximum integration time of 
10 sec to prevent images from being spuriously elongated.
The nod wait was set to $10 - 12$ sec to allow the telescope 
to settle after each nodding, and this resulted in a typical 
duty cycle of about 50\%.

The resulting multiple sets of beam-switched on-chip images 
were individually derotated with respect to each other and 
then co-added into a final image of size $\sim 10\arcsec$ across.
The relative shifts among images were calculated using a 
cross-correlation routine.
Each image was subdivided by $4 \times 4$ pixels
before derotation to make the pixel scale of $0\arcsec.0375$/pix 
for accurate registration.
The total integration times of the final images
are 200 sec for 8.8 and 9.8 $\um$ and 100 sec for 11.7 $\um$,
resulting in 1 $\sigma$ rms noise of 28, 41, and 107 mJy arcsec$^{-2}$
respectively at 8.8, 9.8, and 11.7 $\um$.
For point-spread-function (PSF) and flux calibration,
we observed $\beta$ Peg (a CGS3 standard; \citealt{cohen95})
before and after the object to check for variations in the PSF.
The PSF sizes defined by the full width at half maximum (FWHM)
are 0\arcsec.49, 0\arcsec.51, and 0\arcsec.55 at 8.8, 9.8, and 
11.7 $\um$, respectively.
These values define the observed seeing-limited resolution.
Upon flux calibration, the images were corrected for atmospheric
extinction using the averaged correction factors derived from 
the standard star observations made at various airmasses
throughout the night.
The source flux densities (Jy) at the observed wavelengths were 
derived by scaling the standard flux densities at the corresponding
wavelengths. 
Absolute flux calibration errors are estimated to be approximately 
$\pm 5\%$.
A detailed description of the general MIRAC data reduction and 
flux calibration processes is given by \citet{meixner99}.

We also obtained images of \iras 22272 observed with the 
MIRLIN mid-IR camera \citep{ressler94} mounted at the 
f/40 bent Cassegrain visitor port of the Keck II Telescope. 
The MIRLIN uses a $128 \times 128$ Si:As array with a plate scale 
of 0\arcsec137 pix$^{-1}$ for a total field of view of 
$17\arcsec \times  17\arcsec$. 
The observations were carried out 
on 2000 November 4 under partly cloudy sky.
The object was imaged with an $8\%$ bandwidth filter
at 7.9 $\um$ (N0) and $10\%$ bandwidth filters at 9.7 (N2) 
and 12.5 $\um$ (N5).
The use of a nod-chop beam-switching with 
the N-S nod throw of 7\arcsec~and E-W chop throw of 
7\arcsec~resulted in four on-chip sky-subtracted images of 
the target in a single integration cycle.
The resulting multiple sets of images were derotated 
and co-added into a final image by using the same
routines as the MMT data reduction.
For the purpose of further data analysis, the individual 
on-chip images were subdivided prior to derotation so that 
the final images would have the same pixel scale as the MMT 
images.
The total integration times of the final images
are then 240, 192, and 150 sec respectively for 
7.9, 9.7, and 12.5 $\um$.
$\beta$ And (a CGS3 standard; \citealt{cohen95}) was observed 
before the source observations for the calibration purposes.
However, we were unable to flux calibrate the data due to
the non-photometric sky conditions of the night: we used the 
Keck images only in qualitative analyses in the following.
The PSF sizes are measured to be 0\arcsec.37, 0\arcsec.30,
and 0\arcsec.37 respectively at 7.9, 9.7, and 12.5 $\um$.
Table 1 summarizes the observations along with 
measured quantities to be discussed in the following sections.

\subsection{Sub-arcsecond Mid-IR Morphology}

\subsubsection{Structure of the Mid-IR Nebula}

Images of \iras 22272 and corresponding PSFs
obtained at both observatories are displayed in Figure 1.
The difference in resolution between the telescopes is
readily recognized by the PSF sizes.
The mid-IR emission regions of \iras 22272 consists 
of two major parts: 
the extended halo ($\ale 40 \%$ of the peak intensity) and
emission core ($\age 40 \%$ of the peak intensity).
The emission halo is roughly circular with radius of 
about $1\arcsec$ at the $10 \%$ of the peak intensity 
for all wavebands.
Although not apparent in Figure 1, a faint emission 
halo can be traced out to at least $3\arcsec$ from
the center at the full width at zero intensity.
In order to align images that were obtained at two
observatories with different resolution, 
we first defined the nebula center by taking advantage 
of the circular shape of the emission halo.
We fit the emission halo with elliptical isophotes 
and determined the nebula center by averaging the 
central coordinates of the isophotes with
$20 - 40 \%$ of the peak intensity.
This particular isophotal intensity range was adopted 
to avoid influences from any core structure ($\ale 40 \%$)
while keeping the surface brightness of the halo 
itself reasonably high in order to fit the 
halo shape accurately ($\age 20 \%$):
the $20 \%$ of the peak intensity translated into 
a signal-to-noise (S/N) ratio of at least 50
in these images.
The images presented in Figure 1 have been centered at 
the nebula center.

Within the circular halo, the emission core is 
clearly resolved into a two-peaked structure
in the MMT image at 11.7 $\um$ and 
all three Keck images at 7.9, 9.7, and 12.5 $\um$ 
with the separation of the peaks 
varying from 0\arcsec.44 to 0\arcsec.62.
However, the 8.8 and 9.8 $\um$ MMT images
show merely a single peak.
Irrespective of the detailed structure, 
the emission core shows an overall elongation 
(the averaged ellipticity: 0.14) in the NE-SW direction 
with the position angle (PA) about $55^{\circ}$ (to E from N),
which is consistent with the previous observations at 
arcsecond resolution \citep{meixner97, dayal98}.
The present observations at sub-arcsecond resolution
additionally show a dent at the SE side of the 
elongated core in all wavebands, making the emission 
core ``kinked'' towards the NW direction.
The two emission peaks are located at each end of 
the ``kinked'' core, almost diametrically positioned with 
respect to the nebula center.
The overall appearance of the mid-IR nebula thus 
seems symmetric with respect to a fiducial line with 
a PA of $\sim -35^{\circ}$, which is perpendicular 
to the direction of the general core elongation.

\subsubsection{Structure of the Emission Core}

The two-peaked morphology at the innermost core 
is just as expected from the SOLE-toroidal 
morphological classification of \iras 22272 
\citep{meixner97,ueta00} and agrees well with 
the other two SOLE-toroidal type PPNs, 
\iras 07134+1005 and \iras 17436+5003,
in which the two-peaked core structure was
resolved in the mid-IR (See Figure 5 of 
\citealt{ueta00}).
However, the kinked core shape is unique to 
\iras 22272 and was not observed in the other 
two SOLE-toroidal type PPNs.
While an intermediate inclination angle%
\footnote{The inclination angle of the dust 
torus, $\thincl$, is defined to be the acute 
angle between the line of sight and the pole.}
($\thincl \age 45^{\circ}$) 
was suggested for the central dust torus in 
those two PPNs 
(\citealt{meixner97, meixner00}, respectively), 
a less tilted inclination 
($\thincl = 30^{\circ} - 40^{\circ}$)
was proposed for \iras 22272 based on the 
morphology of a color temperature map \citep{dayal98}.
A smaller $\thincl$ would probably introduce 
such a kinked core morphology due to a difference 
in column densities along the different lines of 
sight in the nebula.
Thus, the mid-IR morphology of \iras 22272 
would indeed be manifestations of the limb-brightened 
edges of the central dust torus 
whose axis of symmetry lies along a PA of about
$-35^{\circ}$ in the plane of the sky.

The emission core structure also shows dramatic 
changes as the waveband shifts.
At 7.9 $\um$, the peak separation is 0\arcsec.44 and
the NE peak is about 40\% brighter than the SW peak.
As one moves to longer wavebands, the separation increases
while relative brightness of the NE peak with respect to
the SW peak decreases.
Especially remarkable is that the SW peak becomes brighter 
than the NE peak at 12.5 $\um$.
These changes are well demonstrated by the normalized 
emission profiles across the direction of the 
general core elongation shown as Figure 2.
The Keck images suggest that the peak separation at 
8.8 and 9.8 $\um$ would be roughly 0\arcsec.45, which 
is smaller than the MMT PSF size at these wavelengths
($\sim 0\arcsec.5$).
This is why the MMT images at 8.8 and 9.8 $\um$ were 
unable to resolve the core structure.
The unresolved nature of the MMT images at lower wavebands 
is also seen in the emission profile
(Figure 2) as the barely resolved profile being 
blended into one structure peaking closer to the
nebula center.
A further inspection of Figure 2 also reveals that 
the location of the NE peak shifts farther away from 
the nebula center at longer wavebands (at 11.7 and 
12.5 $\um$) while the SW peak seems anchored at the 
same position.
This shift is caused primarily by the NE peak that 
increased its brightness at the lower wavebands.

The central star can be bright enough to contribute 
significantly to the total flux at $\ale 10 \um$ in 
the case of an optically thin dust shell.
The proximity of the dominant NE peak to the nebula 
center in the 7.9 and 9.7 $\um$ image is very 
suggestive that the central star is indeed 
responsible for this unique emission characteristics 
of the nebula.
In the optically thin approximation, we can derive 
a color temperature map by taking a ratio of images
at two wavebands (e.g., \citealt{hora96, dayal98}).
Such color temperature maps can be used to infer the 
location of the central star from the distribution of 
the warmest dust grains (e.g., \citealt{ueta01}).
The color temperature maps derived from the MMT images 
showed a single peak between the NE peak and the nebula 
center: these three points (the NE peak, the nebula 
center, and the dust temperature peak) are coincident 
with each other within the resolution of the images.
The color temperature maps derived from the uncalibrated 
Keck images yielded a qualitatively similar result, 
confirming this finding.

\subsubsection{Structure of the Dust Shell}

In order to recover the emission structure by the 
dust shell alone, we removed the central stellar
emission component from the total nebula emission 
by means of 
(1) PSF subtraction and 
(2) deconvolution.
PSF subtraction was done first by estimating the 
stellar flux at the wavelengths of observation from 
the Rayleigh-Jeans tail of the blackbody curve fitted 
with existing near-IR photometry
\citep{manchado89, vhg89, hrivnak91}
including our own%
\footnote{\iras 22272 was observed on 1999 November 16 under
photometric conditions with NIRIM \citep{nirim} at Mt. Laguna 
observatory, which is jointly operated by the University of 
Illinois at Urbana-Champaign and the San Diego State University.
See Table 1.
For more details on data reduction, see \citet{ueta01}.},
and then, by subtracting the scaled PSF images from the 
observed images.
The PSF subtracted images are displayed in the top
frames of Figure 3 (labeled as ``Sub''). 
The central stellar contribution to the total flux 
turned out to be at most $6\%$ (Table 1), and 
the structure of the mid-IR nebula was virtually
unchanged.
We then performed Richardson-Lucy deconvolution to
remove the PSF effects from the ``raw'' images and 
the results are presented in the bottom frames 
of Figure 3 (labeled as ``Decon''). 
The deconvolved 11.7 $\um$ image clearly recovered 
the limb-brightened edges of the dust torus that 
were symmetrically located with respect to 
the relatively emission-free nebula center.

The orientation of the limb-brightened edges of 
the dust torus agrees very well with the direction 
of the suspected axis of symmetry of the dust 
torus along a PA of about $-35^{\circ}$, which 
was also suggested by the elongation of the 
optical and near-IR reflection nebulosity 
\citep{gledhill00,ueta00}.
Figure 4 shows the $V$ band HST image 
\citep{ueta00} overlaid 
with contours of the deconvolved 11.7 $\um$ image.
Aside from the detailed shapes, the directions of
a general elongation in the optical (PA $\sim -40^{\circ}$) 
and mid-IR (PA $\sim 50^{\circ}$) appear to be 
perpendicular, supporting the idea that the 
toroidal dust shell would define the ``waist'' 
of the reflection nebulosity.
Although stellar photons basically can go all directions 
in case of the optically thin dust shell of \iras 22272, 
its dust torus is still capable of shaping the
optical reflection nebulosity according to its
dust density distribution.
The HST image of \iras 22272 shows four elliptical 
tips with the bottom two (along PAs of $\sim 155^{\circ}$ 
and $-75^{\circ}$) being more extensive than the other two.
Interestingly, the directions of these optical 
protrusions are strikingly coincident with the 
directions in which there is less amount 
of dust grains:
the largest elliptical protrusion towards SE
is coincident with one of the bicone openings 
for the dust torus
and other protrusions towards W and N seem 
to correspond with the ``breaks'' of the
dust distribution as indicated by the arrows
in Figure 4.
These consistencies with images at other wavelengths 
strongly suggest that the dust shell assumes
a toroidal shape and its axis of symmetry is oriented 
along PA of $\sim 155^{\circ}$ .

\subsubsection{Non-Stellar Emission Component}

On the contrary, neither PSF subtraction nor
deconvolution successfully removed the
emission component at the nebula center
at 8.8 and 9.8 $\um$.
Instead, the deconvolved images at these
wavebands even indicated the dominance of 
the central emission in the entire mid-IR 
emission regions.
Thus, we deconvolved the PSF subtracted images
at these wavebands.
The deconvolved, PSF-subtracted image at 9.8 $\um$
recovered the dust shell structure that was free
of the central emission.
This would indicate that the PSF subtraction had 
already removed the stellar emission component 
successfully.
However, the deconvolved, PSF subtracted image 
at 8.8 $\um$ still displayed a dominant central 
peak.
These results could suggest that the remaining 
central emission after both PSF subtraction and 
deconvolution represents an emission component 
of non-stellar origin,
that is, there is distribution of dust grains
in the inner cavity of the dust shell.
The existence of such an emission component would 
require post-AGB mass ejection processes.
In fact, such a post-AGB mass loss has been already 
suggested by the observed change of the near-IR CO 
emission features into absorption features,
indicating the onset of a sudden mass ejection
\citep{hkg94}.

\section{2-D Radiative Transfer Modeling}

\subsection{Basic Description}

The mid-IR images at sub-arcsecond resolution revealed 
evidence for the toroidal dust distribution at the
inner edge of the PPN shell of \iras 22272.
The images also suggested a rather small 
inclination angle of the shell 
($\thincl = 30^{\circ} - 40^{\circ}$) and a possible 
presence of post-AGB ejecta in the inner cavity
of the PPN shell.
We can quantify the axisymmetric nature of the 
dust shell with the help of the spatial information 
yielded from these high-resolution mid-IR images.
To achieve this goal, we performed radiative transfer 
calculations with a code which solves the equation 
of radiative transfer in a fully two-dimensional grid 
using a method developed by \citet{collison91}.
This code has been used to model axisymmetric 
circumstellar dust shells around evolved stars
\citep{meixner97,skinner97,ueta01,meixner00}.
In particular, \iras 22272 was previously modeled
with an earlier version of the code using mid-IR 
images at arcsecond resolution showing only the 
unresolved core.
The model calculations predicted the inner shell 
radius ($R_{\rm in}$) to be $0\arcsec.6$ \citep{meixner97}
despite the lack of detailed spatial information.
Interested readers are encouraged to refer to 
\citet{meixner97} and \citet{skinner97} for the detailed 
description of the code and \citet{ueta01} for more 
discussions on the iteration and parameter fitting processes.

Following \citet{meixner97}, the circumstellar dust 
shell in our model calculations is assumed to be a result 
of a two-phased AGB mass loss process:
a spherically symmetric AGB wind (phase 1) gradually 
transforming into an axisymmetric superwind (phase 2)
in the late AGB phase.
In this model, an axisymmetric superwind shell (i.e., 
the central torus) of size $R_{\rm sw}$ is surrounded
by a spherical AGB shell of size $R_{\rm out}$, and
the degree of axisymmetry for the entire dust shell is 
controlled by five geometric parameters in the density 
profile, $\rho (r,\theta)$, and the opening angle of 
the biconical cavities ($\theta_{0}$), in which dust 
density can be set arbitrarily.
The main physical processes, i.e., dust absorption, 
emission, and (isotropic) scattering, are treated at 
zone centers in the 2-D grid by calculating the 
optical properties for a given dust composition 
evaluated with Mie theory.
To account for the near-IR hydrocarbon features seen 
in the spectrum of \iras 22272 \citep{buss90, geballe92, 
justtanont96},
we adopted the optical constants of hydrogenated 
amorphous carbons (HACs; type BE of \citet{colangeli95}
and \citet{zubko96})
and aimed at fitting the overall shape of the continuum 
emission in the SED.
The use of HACs to simulate dust continuum emission in 
\iras 22272 is warranted because \iras 22272 is
known to show especially strong 3.4 and 6.9 $\um$ features 
which arise from the $-$CH$_{2,3}$ functional groups 
(e.g., \citealt{buss93,duley00}).

Starting with the initial parameters adopted from 
the previous calculations and other references in 
the literature (\citealt{meixner97} and 
references therein),
the best-fit model parameters are sought iteratively 
by fitting the shape of the SED as well as the projected 
2-D morphology of the model.
The major differences between our model calculations 
and the previous ones by \citet{meixner97} are that
(1) $R_{\rm in}$ is practically an input 
parameter provided by the images with the resolved 
core structure (Figure 1 and 3),
(2) the code now allows a distribution of grain sizes of
the form $n(a) \propto a^{-3.5} \exp (-a/a_{0})$ 
\citep{kim94,jura94}, 
in which $a$ is the grain size and $a_{0}$ is an exponential 
scaling factor acting as the ``effective'' maximum grain size, and
(3) when applicable (i.e., $2 \pi a / \lambda \ll 1$), the 
assumption of the continuum distribution of ellipsoids 
\citep{bohren83} is used for the particle shape distribution.
Especially, having $R_{\rm in}$ as an input parameter 
is the major strength in our model calculations.
The temperature of the dust shell, $T_{\rm dust}$, 
is essentially defined by $R_{\rm in}$, and therefore, 
the energetics of the dust shell heating is constrained 
fairly well in our model calculations compared 
with other radiative transfer models in which 
$R_{\rm in}$ needs to be iteratively searched by 
fitting only with the shape of the SED.
We adopted 0\arcsec.5 as the initial $R_{\rm in}$
from the deconvolved images (Figure 3).
The use of the high-resolution mid-IR images in the 
model fitting process is robust because the geometrical 
parameters of the dust shell, especially 
the inclination angle ($\thincl$) and biconical 
opening angle ($\theta_{0}$), are rather
sensitive to the morphology of the 2-D projected 
model images.
The dust size distribution also plays an important 
role in determining the best-fit parameters because
$R_{\rm in}$ is now observationally constrained
and the energetics of dust heating can be affected
by different size distributions. 

\subsection{The Best-Fit Model}

The best-fit model for the dust shell of \iras 22272 
consists of the central star ($T_{\rm eff} \sim 5800$ K) 
surrounded by two separate sets of dust shells representing
a PPN shell (AGB wind shell + superwind shell) and
a post-AGB wind shell located in the inner cavity of
the PPN shell.
The PPN shell has the inner radius of 
$R_{\rm in} = 0\arcsec.5$ and consists of the inner,
superwind shell of radius, 
$R_{\rm sw} = 5 \times R_{\rm in}$,
and the outer, AGB wind shell of radius, 
$R_{\rm out} = 24 \times R_{\rm in}$.
The superwind shell is highly 
equatorially-enhanced 
($\rho_{\rm eq}/\rho_{\rm pole} = 9$) as
opposed to the generally spherical AGB wind shell.
The total PPN shell size ($R_{\rm out}$) is
strictly an assumption based on the known
extent of $^{12}$CO emission (e.g., \citealt{fong00})
and needs to be constrained
by future far-IR and/or sub-mm observations.
Within the inner cavity of the PPN shell,
there is another set of shells which account for a 
possible presence of the post-AGB ejecta.
For simplicity, 
the post-AGB shell is assumed to have a similar 
two-layer structure as the PPN shell of the
radius $R_{\rm in}$ (i.e., filling the inner cavity of
the PPN shell), in which the majority of ejecta is 
thought to be concentrated within a thin shell 
defined by the boundaries at 
$R_{\rm in}^{\rm pAGB} = 0\arcsec.075$ and 
$R_{\rm out}^{\rm pAGB} = 1.2 \times R_{\rm in}^{\rm pAGB}$.
The dust size distribution is slightly different in the
superwind shell and the AGB wind shell, i.e., a 
population of large particles is weighted more in the 
superwind shell 
($a_0 = 10$ $\um$) than in the AGB wind shell 
($a_0 = 0.1$ $\um$).
In the post-AGB shell, the dust size distribution
is set to be the same as in the superwind shell
for simplicity.

Because both the luminosity of the central star and the dust 
shell size scale with the square of distance to the source, 
model calculations would not be able to fix the distance
by the iterative fitting.
Therefore, we adopted 1.6 kpc as the distance to \iras 22272
from a recent model calculations \citep{szczerba97} and
estimates (\citealt{yuasa99}; J. Nakashima 2000 priv.\/ comm.).
The adopted distance yields 
$L_{*} \sim 1.3 \times 10^{4} L_{\odot}$ for the 
luminosity of the central star,
$R_{\rm in} = 1.2 \times 10^{16}$ cm,
$R_{\rm sw} = 6.0 \times 10^{16}$ cm, and
$R_{\rm out} = 2.9 \times 10^{17}$ cm for
the PPN shell, and
$R_{\rm in}^{\rm pAGB} = 1.8 \times 10^{15}$ cm, and
$R_{\rm out}^{\rm pAGB} = 2.1 \times 10^{15}$ cm for
the post-AGB shell.
These shell sizes would suggest the dynamical age of 
the shell to be about 380 years,
assuming the constant expansion velocity of 10 km s$^{-1}$ 
\citep{zd86,lind88,woodsworth90,omont93},
and
the AGB and superwind mass loss rates as
${\dot M}_{\rm AGB} = 7.8 \times 10^{-7}$ and 
${\dot M}_{\rm sw} = 4.1 \times 10^{-6} M_{\odot}$ yr$^{-1}$
and the post-AGB mass loss rate as
$2 - 6 \times 10^{-7} M_{\odot}$ yr$^{-1}$,
assuming a dust-to-gas mass ratio of 
$4.5 \times 10^{-3}$, an approximate value used for
C$-$rich AGB stars \citep{jura86}.
Table 2 summarizes the parameters and 
derived quantities for the best-fit model.

Figure 5 shows the best-fit SED for our model calculations.
The total SED consists of the PPN shell contribution and 
the post-AGB shell contribution.
The {\sl Infrared Space Observatory} ({\sl ISO}) 
spectrum%
\footnote{The spectrum is constructed by 
combining the pipeline-calibrated SWS and LWS data 
available at the {\sl ISO\/} archive 
(http://www.iso.vilspa.esa.es/ida/index\_us.html).} 
and other photometric data
(\citealt{manchado89,vhg89,hrivnak91,lario97,meixner97,dayal98},
\iras fluxes, and present observations)
are shown for comparison.
As the {\sl ISO\/} spectrum shows, \iras 22272 is
full of strong IR features on the blueward shoulder 
and near the dust peak 
(the far-IR spikes appearing in the LWS data at 
$\sim 200$ $\um$ seem to be instrumental), 
and it is difficult to define the continuum:
we assumed fluxes at around 15 and 60 $\um$ represented
the continuum.
The $1.3$ mm continuum flux is strictly a lower limit 
because the observed beam size (11\arcsec) is smaller 
than the known extent of the molecular envelope 
(e.g., $^{12}$CO shell of $18\arcsec$ diameter; 
\citealt{fong00}).
\iras 22272 is visually known as a variable star and 
the variation is indicated by the optical photometry 
done over two epochs \citep{hrivnak91}, each of which 
is connected by a thin line.
The interstellar extinction has been taken
into account and the best-fit model suggests
$A_{\rm v} = 2.5$, which is slightly higher than 
the observed value along the line of sight 
toward \iras 22272 for the adopted distance,
1.6 kpc \citep{neckel80}.

Figure 6 shows the 2-D projected images for our 
best-fit models at 8.8, 9.7, and 11.7 $\um$.
To create these images, we convolved only the PPN 
shell model images with the observed PSF images.
By inspection, one immediately sees that emission 
at the nebula center is lacking in model images
at 8.8 and 9.8 $\um$.
This is due to the fact that the post-AGB shell 
model was not included, and is consistent with the 
inference from the observations that emission
near the nebula center at lower wavebands arise
from the post-AGB ejecta located in the inner cavity 
of the PPN shell.
The whole shell system has biconical openings of
$20^{\circ}$ along the axis of symmetry 
with an inclination of $\sim 25^{\circ}$ 
from the pole-on orientation, and
the near side of the biconical openings points 
to the SE direction (PA of $150^{\circ}$).
The resulting morphologies become extremely sensitive 
to the $\thincl$ and $\theta_{0}$ pair when the dust 
shell is optically thin as \iras 22272 and 
would completely be altered if, for example, 
$\thincl$ is changed by more than $\sim 10\%$.

\subsection{The Energetics of Dust Shell Heating}

For our model calculations,
we first adopted the dust size distribution,
$n(a) \propto a^{-3.5} \exp (-a/a_{0})$, with 
$a_{0} = 0.1$ $\um$ for the entire PPN shell,
based on the study of the particle size distribution 
around a well-known carbon star, IRC+10 216 \citep{jura94}.
This dust distribution, however, yielded the SED whose 
dust peak was too blue, i.e., $T_{\rm dust}$ was too high.
This SED is shown in Figure 4 as a thin dotted line
for comparison.
In typical radiative transfer calculations, $R_{\rm in}$ 
would then be iteratively increased to decrease 
$T_{\rm dust}$ (i.e., to shift the dust peak redward).
Because $R_{\rm in}$ was already well-constrained 
to be around 0\arcsec.5 by the mid-IR images, 
increasing $R_{\rm in}$ would not be a desirable option 
in our iterations.
In Mie theory, the scattering and absorption coefficients 
are respectively proportional to $(a/\lambda)^{4}$ and 
$a/\lambda$ if $a/\lambda$ is sufficiently small,
that is, smaller particles are generally more absorptive
than larger particles.
Therefore, increasing $a_0$ is physically equivalent to 
decreasing the efficiency of dust absorption, and 
the energetics of dust shell heating can be 
controlled by adjusting the efficiency of dust
absorption, i.e., $a_0$.

We then iteratively changed $a_0$ in both of the 
superwind and AGB wind shells and converged to the 
best-fit model (thick dashed line), in which the 
superwind shell has a 
larger population of big grains ($a_0 = 10$ $\um$) than 
the AGB wind shell ($a_0 = 0.1$ $\um$) 
while $a_{\rm min}$ is kept at the
best-fit value of 10 $\AA$.
Upon iterating on the size distribution, we simply 
assumed that any change of size distribution would
concurrently occur with the change of the mass loss
geometry.
The superwind shell is thus made slightly inefficient 
in absorbing the stellar UV/optical photons, effectively 
reducing $T_{\rm dust}$ at $R_{\rm in}$.
In the AGB wind shell, there is now more abundant
UV/optical photons that have been scattered through 
the superwind shell,
and this excess population of UV/optical photons 
contributes to an additional heating of the AGB 
wind shell increasing the total far-IR flux.
The population of small particles in the AGB wind 
shell is thus necessary as the absorbers of 
stray UV/optical photons to be an additional source
for the dust shell heating.
If $a_0$ is set to 10 $\um$ in the AGB wind shell, 
for example, the dust heating would become too 
inefficient and there would not be enough IR flux 
to fit both of the blueward shoulder and redward 
tail of the observed dust peak.

One would intuitively think that larger grains are
typically found in the inner region of the dust shell, 
where density is highest, and that such large grains 
would be sputtered into smaller pieces as they
coast away from the central star.
Such an intuitive view was theoretically confirmed
by hydrodynamical calculations of a self-consistent 
dust-driven wind model for a C$-$rich star \citep{dominik89}.
Furthermore, \citet{kruger97} found that grain 
drift significantly reduce the grain distribution
at large grain sizes, and the effect was found 
to be more pronounced in winds with lower mass 
loss rates.
These results are consistent with our best-fit model
in which an enhanced population of large dust 
grains is found in the superwind shell, where 
dust density and mass loss rates are higher.

\subsection{The Post-AGB Shell}

The observed dust peak is enormous due to 
the strong IR features that are seen between
$20 - 30$ $\um$ in the {\sl ISO\/} spectrum. 
If we fit the shape of the dust peak using 
the assumed continuum at around 15 and 60 $\um$,
the PPN shell model alone would significantly 
underestimate fluxes at these wavelengths 
(see the difference between the {\sl ISO\/} 
spectrum and thick dashed line at around $5 - 10$ $\um$).
Although some flux underestimate in this region of 
the spectrum is expected due to lack of proper 
identification of the dust feature carriers,
the resulting flux underestimate seems too severe 
to be accounted for just by the insufficient
knowledge of dust composition in the PPN shell.

Therefore, we also performed radiative transfer 
calculations for a post-AGB shell - distribution of
mid-IR emitting material located in the inner cavity
of the PPN shell - as suggested by the centrally 
concentrated core morphology 
observed in the lower waveband images.
The ultra-sub-arcsecond structure of the post-AGB 
shell can not be determined from our mid-IR images 
and the nature of a post-AGB mass ejection is 
generally not well understood.
To date, there is only one observational suggestion
for the post-AGB mass ejection in \iras 22272,
in which the CO emission features transformed 
into absorption features about a decade ago
\citep{hkg94}.
Thus, we simply assumed that an ambient post-AGB mass 
loss continued since the termination of the
superwind and a sudden mass ejection took place
about 10 years ago.
The width of the post-AGB shell is then determined by
assuming any mass loss would result in an expansion
at the rate of 10 km s$^{-1}$
(observed CO expansion velocity; \citealt{woodsworth90}).
The inner post-AGB shell radius ($R_{\rm in}^{\rm pAGB}$) 
is estimated through iterations so that the SED, 
$T_{\rm dust}$ at $R_{\rm in}^{\rm pAGB}$, and
projected 2-D image of the post-AGB shell would be 
consistent with the observations.

The best-fit SED for the post-AGB shell 
(thick dashed-dotted line) was then combined 
with the best-fit SED for the AGB wind shell 
to construct the total SED for the model.
These two best-fit models were derived through independent
model calculations because the code is not capable of 
treating two shells together, and thus the total model 
is not strictly self-consistent in a sense that the
AGB wind shell does not ``know'' the presence of a post-AGB
shell within its inner cavity.
However, the post-AGB shell is so optically thin that
the input stellar SED (thick dotted line) is virtually 
unaffected by the presence of the post-AGB shell
and thus we conclude that the AGB wind shell model 
is valid even when the post-AGB wind shell is present.

\section{Discussions}

\subsection{Asymmetric Appearance of the Dust Torus}

Our identification of the toroidal dust distribution
in the PPN shell is based on the morphology of the 
mid-IR emission core: there should be two emission 
peaks, which represent the limb-brightened edges of 
a dust torus, oriented diametrically symmetric with 
respect to the nebula center.
In reality, however, the mid-IR nebula usually appear
somewhat asymmetric because one peak is brighter and/or 
more extended than the other.
The 11.7 $\um$ image (bottom left in Figure 1) 
exemplifies such an asymmetry as the imbalance of the
peak brightness and size.
These asymmetries can be attributed to 
physical asymmetries of the dust torus,
such as asymmetric mass loss
\citep{jura01},
and an inhomogeneity in the dust distribution
\citep{dayal98}.
It may, however, simply be because of the PSF effect.

Despite the asymmetric appearance of the raw image,
the deconvolved 11.7 $\um$ image recovered a highly
symmetric morphology of the limb-brightened edges.
Comparison with an HST image also shows that the 
limb-brightened peaks are equidistantly positioned
with respect to the central star (Figure 3).
Meanwhile, 
a 2-D projected image of the PPN shell model,
which is completely axisymmetric, yielded an
asymmetric morphology similar to the observed
one after being convolved with the observed PSF image.
Thus, no physical asymmetry is required to
explain an asymmetric appearance seen in the 
mid-IR images of the PPN shells, and
we argue that our axisymmetric shell model
would represent the real PPN shells 
sufficiently well.
This is also supported by the fact that the Keck 
12.5 $\um$ image looks more axisymmetric than the MMT
11.7 $\um$ image (bottom frames in Figure 1),
which is due to the difference in the PSF shapes:
the Keck PSFs appear more symmetric than the
MMT PSFs.
It is, however, of interest to note that the
emission profiles shown in Figure 2 indicates
that the NE peak is brighter at lower
wavelengths than the SW peak, i.e., the
NE peak is warmer than the SW peak.

The origins of axisymmetry in the PPN dust shells,
however, are still unclear in spite of
a number of possible mechanisms that have been
suggested.
Such ideas includes the binary interaction 
(e.g., \citealt{mm99}),
the magneto-hydrodynamic effects 
(e.g., \citealt{segura00}), and
the remnant planetary systems or
re-forming accretion disks
(e.g., \citealt{soker97}).
Whichever the true scenario may be, the 
axisymmetry generating mechanisms have to be able to 
create a high equatorial enhancement 
($\rho_{\rm eq}/\rho_{\rm pole} = 9$)
in the density distribution of the PPN dust shells.

\subsection{Dust Size Distribution in the PPN Shell}

The resolved mid-IR images placed a rather strict
constraint on $R_{\rm in}$.
With the energy output ($L_*$) and  $\teff$ of the 
central star also fairly well-constrained by observations
\citep{kvh89,zacs95},
$T_{\rm dust}$ is rather tightly fixed unless
grain properties are modified.
In order to fit the SED, we iterated on the dust size 
distribution to adjust the energetics in the PPN shell,
and the best-fit model suggested the existence of two 
distinct size distributions:
there is a population of larger grains in the superwind 
shell than the AGB shell.
In general, large dust grains are required to cause
polarization \citep{jura94}, and the best-fit model 
seems to suggest that one would observe polarized
emission from the superwind shell of \iras 22272.
\citet{gledhill00} imaged \iras 22272 by means of imaging 
polarimetry and discovered that the polarized emission 
mainly arose from the elongated shell of dust
surrounding the central star.
According to their $J$ band image of \iras 22272
(Figure 15 of \citealt{gledhill00}),
the extent of the elongated polarized emission shell 
is  $4\arcsec \times 2\arcsec.9$ with 
$R_{\rm in} = 0\arcsec.7$.
Our model and the results from the imaging polarimetry
agree very well and strongly suggest the presence of
a population of larger dust grains in the superwind 
shell of \iras 22272.

Radiative transfer calculations by \citet{szczerba97}
yielded $R_{\rm in} = 0\arcsec.94$, which is about 
twice as large as our observed/best-fit value. 
This difference probably originates mostly from 
the difference in the size distribution used in 
the models.
Their dust distribution includes a significantly 
smaller population of grains compared with ours 
with the absolute minimum and maximum sizes of 
$5\AA$ and 0.25 $\um$.
Therefore, their model calculations would have yielded 
too high $T_{\rm dust}$ and too much IR excess with the 
observed value of $R_{\rm in} = 0\arcsec.5$ because 
of their use of smaller, more absorption-efficient
dust grains, even if we take into account the fact that their 
spherically symmetric shell structure would provide 
more dust grains to heat at $R_{\rm in}$.

The derived AGB and superwind mass loss rates 
($8 - 40 \times 10^{-7} M_{\odot}$ yr$^{-1}$) 
are about a factor of 2 lower than the previously 
estimated values from $^{12}$CO observations 
($9 - 90 \times 10^{-7} M_{\odot}$ yr$^{-1}$;
\citealt{omont93}).
As seen in the model SED, our calculations underestimated
the IR excess at around $20 - 40$ $\um$ due to unknown 
dust feature carrier species.
Such underestimate of the IR excess is directly related
to underestimate of dust mass, hence mass loss rates.
Therefore, we would consider, also taking into account 
the uncertainties involved in estimating the mass loss 
rates from CO observations, that our results are consistent
with observations.
The previous model calculations \citep{meixner97}
suggested mass loss rates that were about a factor of
10 higher than ours.
This is primarily because of the inclusion of dust size
distribution in our present model calculations.
The previous calculations used a single dust size
while the present model took into account
particles of larger size.
This means that the dust shell in the previous model
was relatively inefficient in absorbing stellar radiation 
(i.e., smaller cross sections)
and that it took more dust grains to produce
the observed IR excess, subsequently resulting in 
higher mass loss rates.
The derived mass loss rates in the present study are 
reasonable considering the optically {\sl thin}
nature of the dust shell.
It also shows that a proper consideration of the dust size 
distribution is required in radiative transfer
calculations of the PPN dust shells.

In our model, we simply assumed that the change of the 
dust size distribution would concurrently take place 
with the change of mass loss geometry from spherical 
to axial symmetry.
Thus, the choice of $R_{\rm sw}$ and $a_0$ for the 
superwind dust shell is crucial in the model fitting.
Because $R_{\rm sw}$ is rather difficult to estimate,
we need to constrain $a_0$ through 
further photometric observations
of far-IR, sub-mm, and radio continuum fluxes.
Radio continuum fluxes are especially useful because
they would yield the power-law like dust emissivity 
in the Rayleigh-Jeans limit \citep{knapp94},
from which we can infer $a_0$ by a simple analysis
(e.g., \citealt{jura00}).

\subsection{The Post-AGB Mass Loss}

The mid-IR images indicated that there is more centrally 
concentrated emission in the lower waveband images that 
could not be accounted for only by the central star 
(Figures 2 and 3).
The radiative transfer calculations showed 
that, if only the PPN shell was considered, there would 
be a significant flux underestimate
between 5 and 10 $\um$ in the SED (Figure 4) and
lack of emission near the nebula center in the 
lower waveband model images (Figure 5).
Our attempt to boost flux at $5-10$ $\um$ by the
addition of the post-AGB shell yielded an adequate
fit to the observations.
The presence of the post-AGB ejecta in the inner cavity 
of the PPN shell is not entirely {\sl ad hoc} because 
recent episodes of mass loss have been observationally
suggested \citep{hkg94} and such sporadic mass
loss events may be enough to create hot dust grains
within the inner radius of the PPN shell.
\citet{szczerba97} also introduced a hot dust component 
in their radiative transfer models to increase flux at 
$5-10$ $\um$.

These findings all agree to suggest that there seems 
to be a distribution of dust grains located rather 
close to the central star.
The origins of the hot dust grains are not at all 
clear: these grains may be a result of a sudden
mass ejection \citep{hkg94} or may simply be
a dwindling mass loss continued long after the end
of the AGB phase \citep{szczerba97}.
Our model assumes a scenario in which an ambient
mass loss kept continuing after the end of the 
superwind phase at a rate of 
$1.6 \times 10^{-7} M_{\odot}$ yr$^{-1}$
(almost equivalent to ${\dot M}_{\rm AGB}$)
to be overpowered by a sudden mass ejection
presumably occurred 10 years ago at a rate of
${\dot M}_{\rm pAGB} = 5.8 \times 10^{-7} M_{\odot}$ yr$^{-1}$.
These values are comparable to the AGB and superwind
mass loss rates and seem to be too high for
an ``ambient'' post-AGB mass loss.
These estimates are, however, strongly dependent on 
$R_{\rm in}^{\rm pAGB}$, 
the geometry of the post-AGB mass loss,
and the dust size distribution in the post-AGB shell.
In our model, we simply assumed the same dust 
size distribution and the shell geometry as in the 
superwind shell.
If we used, for example, a dust size distribution with
smaller grains and a spherically symmetric shell
geometry, we would needed less amount of 
post-AGB ejecta to produce the unaccounted IR 
excess by the PPN shell.
This effectively reduces the post-AGB mass loss rates.
However, we would not be able constrain these 
parameters with the currently available observational data.
Nevertheless,
the existence of some mid-IR emitting material
in the inner cavity of the PPN shell seems to require 
mass loss processes after the end of the AGB phase.

\section{Conclusions}

We have obtained sub-arcsecond resolution mid-IR images 
of \iras 22272 at 8.8, 9.8, and 11.7 $\um$ using 
MIRAC3/BLINC at the MMT observatory
and at 7.9, 9.7, and 12.5 $\um$ using MIRLIN at the Keck
observatory.
The mid-IR core structure was resolved in the 11.7 $\um$ 
MMT image and at all three waveband Keck images
as two emission peaks directly indicating the presence of 
the central dust torus.
The subsequent image analyses suggested that the PPN shell
of $R_{\rm in} = 0\arcsec.5$ had a rather small inclination
angle from the pole-on position.

The 2-D radiative transfer calculations 
indicated that the mid-IR morphology was adequately
reproduced by an axisymmetric dust shell which 
consists of the superwind shell of $6.0 \times 10^{16}$ cm
and the AGB wind shell of $2.9 \times 10^{17}$ cm,
being inclined by $25^{\circ}$ from the pole-on orientation.
This fairly equatorially-enhanced
($\rho_{\rm eq}/\rho_{\rm pole} = 9$) dust shell 
is assumed to be a product of the largely spherical
AGB wind mass loss of 
${\dot M}_{\rm AGB} = 7.8 \times 10^{-7} M_{\odot}$ yr$^{-1}$ 
followed by
the intrinsically axisymmetric superwind of 
${\dot M}_{\rm sw} = 4.1 \times 10^{-6} M_{\odot}$ yr$^{-1}$,
which was terminated about 380 years ago.

The images at lower wavelengths showed very pronounced emission
at the nebula center, which the model for the PPN shell
failed to reproduce.
The best-fit model also indicated the presence of dust
gains in the inner cavity of the PPN shell which would
seem to have been created after the end of the AGB phase.
Our model, based on the observed signature for a sudden 
mass ejection, suggested that
the post-AGB shell would have 
$R_{\rm in}^{\rm pAGB} = 1.8 \times 10^{15}$ cm
after experiencing an ambient mass loss
at $1.6 \times 10^{-7} M_{\odot}$ yr$^{-1}$
followed by a post-AGB mass ejection at
${\dot M}_{\rm pAGB} = 5.8 \times 10^{-7} M_{\odot}$ yr$^{-1}$,
which took place about 10 years ago.

The resolved images showed asymmetry in the brightness/size
of the two limb-brightened edges of the dust torus.
Our image analyses suggest that the PSF effect would be
the major factor in causing the observed asymmetry
in the mid-IR morphology and no physical asymmetry would
be required in the structure of the PPN shell.
Nevertheless, the axisymmetric shaping of the PPN shell 
still requires mechanisms that are capable of inducing 
a high equator-to-pole density ratio.
A population of large particles within the superwind 
shell was also suggested from the energetics of the dust 
shell heating in the model calculations.
Although our best-fit 2-D model calculations reproduce
a number of observed characteristics,
proper identification of all the dust features 
must be accomplished to further improve the quantitative
analyses.

\acknowledgments

We would like to thank 
the staff at the MMT for their assistance in obtaining
the MMT/MIRAC3 data and
Karl Stapelfeldt, Ralph Neuh\"auser, and
Dave Cole for their help in obtaining the Keck/MIRLIN data.
We also thank Angela K. Speck for discussions on
dust properties and
Jun-ichi Nakashima for discussions on a variety of
distance determination schemes for the evolved stars.
An anonymous referee is also thanked for comments and 
suggestions.
Ueta and Meixner are supported by NSF CAREER Award AST-9733697.
MIRAC upgrade and operation are supported by NSF Grant AST-9618850.
BLINC construction and operation are supported by the Terrestrial
Planet Finder Mission development at NASA's Jet Propulsion 
Laboratory.
This research has made use of 
the SIMBAD database, operated at Centre de Donn{\'e}es astronomiques, 
Strasburg, France, and 
the IRAF data reduction and analysis system,
which is distributed by the National Optical Astronomy 
Observatories operated by the Association of Universities 
for Research in Astronomy, Inc., under cooperative agreement 
with the National Science Foundation.

\clearpage
\figcaption{Observed grayscale images of \iras 22272+5435 at 
8.8, 9.8, and 11.7 $\um$ obtained at the MMT Observatory
(left frames,  labeled as ``MMT'') and at 7.9, 9.7, and 12.5 
$\um$ obtained at the Keck Observatory (right frames;
labeled as ``Keck'') with north 
up and east to the left.
The tick marks show relative offsets in arcseconds.
Contours are spaced by $10\%$ of the peak intensity 
with the innermost contour being $90\%$ of the peak.
The insets show the corresponding PSF standard stars
($\beta$ Peg for the MMT observation and 
$\beta$ And for the Keck observations) in the same
format.
Local peak intensities are
15 Jy arcsec$^{-2}$ at 8.8 $\um$,
14 Jy arcsec$^{-2}$ at 9.8 $\um$, and 
30 and 29 Jy arcsec$^{-2}$ at 11.7 $\um$.
The Keck images were not flux calibrated
due to unphotometric sky conditions.}

\figcaption{Normalized surface intensity profiles of the 
images sliced along a PA of $55^{\circ}$ through the nebula 
center.
The PSF profiles at 7.9 and 8.8 $\um$ are also plotted
to show the extension in the peaks.}

\figcaption{Stellar emission subtracted and Richardson-Lucy
deconvolved images at 
8.8 (left), 9.8 (middle), and 11.7 $\um$ (right).
The displaying conventions follow those of Figure 1.}

\figcaption{The $V$ band {\sl HST} image overlaid with
the 11.7 $\um$ contours following the display convention of Figure 1.
Arrows indicate the directions of the elliptical
protrusions that are coincident with the ``kinked'' bicone
openings defined by the central dust torus.}

\figcaption{The spectral energy distribution of the best-fit
model: the entire shell system (a thick solid line)
consists of the AGB wind shell (thick dashed line)
and the post-AGB wind shell (thick dot-dashed line).
Photometric data are indicated by a thin dot-dashed line
(the {\sl ISO} data) and crosses (various photometric
data).
Also shown (thin dotted line) is a single dust distribution
model having the AGB wind shell dust size distribution 
($a_0 = 0.1$ $\um$) throughout the PPN shell to visualize 
the peak shift incurred by the change of the dust size
distribution. 
The input stellar SED (thick dotted line) is also displayed 
to be compared
with the stellar peak of the post-AGB wind shell model.
In the total model, the best-fit interstellar extinction 
($A_{\rm v} = 2.5$) has been applied.}

\figcaption{The 2-D projected images of 
the PPN shell model (without the post-AGB wind shell)
at 8.8, 9.8, and 11.7 $\um$
following the display convention of Figure 1.}

\begin{deluxetable}{cccccccccccc}
\tabletypesize{\scriptsize}
\tablecolumns{12} 
\tablewidth{0pc} 
\tablecaption{Summary of Infrared Observations of IRAS 22272+5435} 
\tablehead{ 
\colhead{Date} &
\colhead{Camera/Telescope} &
\colhead{$\lambda$} &
\colhead{Size\tablenotemark{1}} &
\colhead{PA\tablenotemark{2}} &
\colhead{PSF Size\tablenotemark{3}} &
\colhead{Flux\tablenotemark{4}} &
\colhead{Peak} &
\colhead{$F_{\rm star}/F_{\rm total}$\tablenotemark{5}} \\
\colhead{} &
\colhead{} &
\colhead{($\um$)} &
\colhead{(arcsec)} &
\colhead{(arcsec)} &
\colhead{(arcsec)} &
\colhead{(Jy or mag)} &
\colhead{(Jy arcsec$^{-2}$)} &
\colhead{}}
\startdata 
2000 Jun 16 & 
MIRAC3/MMT &
\phn8.8 & 
$1.3 \times 1.2$ & 
57 & 
$0.49 \pm 0.06$ &
$26 \pm 1$& 
15 & 
0.06 \\
 &
 &
\phn9.8 & 
$1.4 \times 1.3$ & 
59 & 
$0.51 \pm 0.01$  &
$31 \pm 1$& 
14 & 
0.04 \\
  & 
  &
11.7 & 
$1.7 \times 1.5$ & 
56 & 
$0.55 \pm 0.05$ &
$84 \pm 4$& 
30, 29 & 
0.01 \\
2000 Nov \phn4 &
MIRLIN/Keck II &
\phn7.9 & 
$1.2 \times 0.8$ & 
57 & 
$0.37 \pm 0.02$  &
\nodata & 
\nodata & 
\nodata \\
 &
 &
\phn9.7 & 
$1.4 \times 1.2$ & 
52 & 
$0.30 \pm 0.01$  &
\nodata & 
\nodata &
\nodata \\
 &
 &
12.5 & 
$1.7 \times 1.5$ & 
51 & 
$0.37 \pm 0.01$  &
\nodata & 
\nodata & 
\nodata \\
\hline
1999 Nov 16 &
NIRIM/MLO &
J: 1.257 & 
\nodata & 
\nodata & 
\nodata  &
$5.43 \pm 0.15$& 
\nodata & 
\nodata \\
 &
 &
H: 1.649 & 
\nodata & 
\nodata & 
\nodata &
$4.89 \pm 0.07$ & 
\nodata &
\nodata \\
 &
 &
K': 2.12 & 
\nodata & 
\nodata & 
\nodata  &
$5.10 \pm 0.32$ & 
\nodata & 
\nodata \\
\enddata 
\tablenotetext{1}{Major and minor axis lengths
at $50\%$ of the peak intensity.}
\tablenotetext{2}{Position angle measured counter-clockwise
from the North at $50\%$ of the peak.}
\tablenotetext{3}{FWHM of the standard star.}
\tablenotetext{4}{Jy for the mid-IR observations and mag for the 
near-IR observations.}
\tablenotetext{5}{A flux ratio of the estimated stellar component
to the total dust emission.}
\end{deluxetable}

\begin{deluxetable}{cccc}
\tabletypesize{\scriptsize}
\tablecolumns{4} 
\tablewidth{0pc} 
\tablecaption{Input and Derived Model Quantities} 

\tablehead{
\colhead{} &
\multicolumn{2}{c}{Value} &
\colhead{} \\
\cline{2-3} 
\colhead{Parameters} &
\multicolumn{2}{c}{Central Star} &
\colhead{References}}

\startdata 
$\lstar (\lsun)$ & 
\multicolumn{2}{c}{$1.3 \times 10^{4}$}& 1 \\
$\teff$ (K)  & 
\multicolumn{2}{c}{$5600 \pm 400$} & 2 \\
$d$ (kpc)      & 
\multicolumn{2}{c}{$1.6$} & 3 \\
ISM $A_{\rm V}$     & 
\multicolumn{2}{c}{$2.5 \pm 0.5$}  & 4  \\

\hline
&
\multicolumn{2}{c}{PPN Shell} & \\
\cline{2-3} 
& Superwind Shell & AGB Wind Shell & \\
\hline

Inner Radius (cm) &
$1.2 \times 10^{16}$ ($= R_{\rm in}$) &
$6.0 \times 10^{16}$ ($= R_{\rm sw}$) & 
\\

Outer Radius (cm) &
$6.0 \times 10^{16}$ ($= R_{\rm sw}$) &
$2.9 \times 10^{17}$ ($= R_{\rm out}$) &
\\

${\dot M}$ ($\msun$ yr$^{-1}$) &
$4.1 \times 10^{-6}$ ($= \mdot_{\rm sw}$) &
$7.8 \times 10^{-7}$ ($= \mdot_{\rm AGB}$)&
\\

$M_{\rm dust}$ ($\msun$) &
$2.8 \times 10^{-5}$  &
$2.5 \times 10^{-5}$ &
\\

$T_{\rm dust}$ at $R_{\rm in}$, $R_{\rm sw}$ (K) & 
202 &
84  &  
\\

$\tau_{9.8\um,~{\rm pole}}$   & 
\multicolumn{2}{c}{0.001} & 
\\

$\tau_{9.8\um,~{\rm eq}}$   & 
\multicolumn{2}{c}{0.031} & 
\\

$\theta_{\rm incl}$\tablenotemark{a} & 
\multicolumn{2}{c}{$\phn25 \pm \phn3^{\circ}$} &
\\

$\theta_0$\tablenotemark{a} & 
\multicolumn{2}{c}{$\phn20 \pm \phn5^{\circ}$} &
\\

PA\tablenotemark{a} & 
\multicolumn{2}{c}{$135 \pm 10^{\circ}$} & 
\\

$v_{\rm exp}$\tablenotemark{a} & 
\multicolumn{2}{c}{10 km s$^{-1}$} & 5 \\

Composition &
\multicolumn{2}{c}{Hydrogenated Amorphous Carbons} &
6 \\

Dust Size &
$a > 10 \AA$, $a_0 = 10.0 \um$ &
$a > 10 \AA$, $a_0 = \phn0.1 \um$ & 
\\

\hline
&
\multicolumn{2}{c}{Post-AGB Shell} & \\
\cline{2-3} 
& Sudden Mass Ejection & Ambient Mass Loss & \\
\hline

Inner Radius (cm) &
$1.8 \times 10^{15}$ ($= R_{\rm in}^{\rm pAGB}$) &
$2.1 \times 10^{15}$ ($= R_{\rm out}^{\rm pAGB}$) & 
\\

Outer Radius (cm) &
$2.1 \times 10^{15}$ ($= R_{\rm out}^{\rm pAGB}$) &
$1.2 \times 10^{16}$ ($= R_{\rm in}$) &
\\

${\dot M}$ ($\msun$ yr$^{-1}$) &
$5.8 \times 10^{-7}$ ($= \mdot_{\rm pAGB}$) &
$1.6 \times 10^{-7}$ &
\\

$M_{\rm dust}$ ($\msun$) &
$2.6 \times 10^{-8}$  &
$2.3 \times 10^{-7}$ &
\\

$T_{\rm dust}$ at $R_{\rm in}^{\rm pAGB}$, $R_{\rm out}^{\rm pAGB}$ (K) & 
414 &
385  &  
\\

$\tau_{9.8\um,~{\rm pole}}$   & 
\multicolumn{2}{c}{0.001} & 
\\

$\tau_{9.8\um,~{\rm eq}}$   & 
\multicolumn{2}{c}{0.023} & 
\\

Composition &
\multicolumn{2}{c}{Hydrogenated Amorphous Carbons} &
6 \\

Dust Size &
\multicolumn{2}{c}{$a > 10 \AA$, $a_0 = 10.0 \um$} &
\\

\enddata 
\tablerefs{%
1: \citet{kvh89},
2: \citet{zacs95},
3: \citet{szczerba97,yuasa99}; Nakashima (2000),
4: \citet{neckel80},
5: \citet{zd86,lind88,woodsworth90,omont93},
6: \citet{colangeli95,zubko96}}
\tablenotetext{a}{Same for the post-AGB shell.}
\end{deluxetable}


\begin{thebibliography}{dummy} 

\bibitem[Allamandola, Tielens, \& Barker\/(1989)]{atb89}
  Allamandola, L. J., Tielens, A. G. G. M., \&
  Barker, J. R.
  1989, \apjs, 71, 733

\bibitem[Begeman et al.\/(1996)]{begeman96}
  Begemann, B., Dorschner, J., Henning, Th., \& 
  Mutschke, H.
  1996, \apj, 464, L195

\bibitem[Bohren \& Huffman\/(1983)]{bohren83}
  Bohren, C. F. \& Huffman, D. R.
  1983, Absorption and Scattering of Light by Small Particles
  (New York: John Wiley \& Sons)

\bibitem[Buss et al.\/(1990)]{buss90}
  Buss, R. H., Cohen, M., Tielens, A. G. G. M., Werner, M. W., 
  Bregman, J. D., Witteborn, F. C., Rank, D., \& Sandford, S. A.
  1990, \apj, 365, L23  

\bibitem[Buss et al.\/(1993)]{buss93}
  Buss, R. H., Tielens, A. G. G. M., Cohen, M., Werner, M. W., 
  Bregman, J. D., \& Witteborn, F. C.
  1993, \apj, 415, 250

\bibitem[Cohen \& Davies\/(1995)]{cohen95}
  Cohen, M. \& Davies J. K.
  1995, \mnras, 276, 715

\bibitem[Colangeli et al.\/(1995)]{colangeli95}
  Colangeli, L., Mennella, V., Palumbo, P., Rotundi, A., \&
  Bussoletti, E.
  1995, \aaps, 113, 561

\bibitem[Collison \& Fix\/(1991)]{collison91}
  Collison, A. \& Fix, J.
  1991, \apj, 368, 545

\bibitem[Dayal et al.\/(1998)]{dayal98}
  Dayal, A., Hoffmann, W. F., Bieging, J. H., Hora, J. L., 
  Deutsch, L. K., \& Fazio, G. G. 
  1998, \apj, 492, 603

\bibitem[Dominik, Gail, \& Sedlmayr\/(1989)]{dominik89}
  Dominik, C., Gail, H.-P., \& Sedlmayr, E.
  1989, \aap, 223, 227

\bibitem[Duley\/(2000)]{duley00}
  Duley, W. W.
  2000, \apj, 528, 841

\bibitem[Fong et al.\/(2000)]{fong00}
  Fong, D., Meixner, M., Sutton, E. C., Welch, W. J., Bujarrabal, V.,
  \& Castro-Carrizo, A.
  2000, in Asymmetrical Planetary Nebulae II: From Origins to
  Microstructures, ed. J. H. Kastner, N. Soker, \& S. A. Rappaport
  (San Fransisco: ASP), 87

\bibitem[Garc{\' \i}a-Lario et al.\/(1997)]{lario97}
  Garc{\' \i}a-Lario, P., Manchado, A., Pych, W., \& 
  Pottasch, S. R.
  1997, \aaps, 126, 479

\bibitem[Garc{\' \i}a-Segura et al.\/(2000)]{segura00}
  Garc{\' \i}a-Segura, G., Franco, J., L\'opez, J. A.,
  Langer, N., \& R\'o\.zyczka, M.
  2000, in Asymmetrical Planetary Nebulae II: From Origins to
  Microstructures, ed. J. H. Kastner, N. Soker, \& S. A. Rappaport
  (San Fransisco: ASP), 234
  
\bibitem[Geballe et al.\/(1992)]{geballe92}
  Geballe, T. R., Tielens, A. G. G. M., Kwok, S., \& Hrivnak, B. J.
  1992, \apj, 387, L89

\bibitem[Gledhill et al.\/(2000)]{gledhill00}
  Gledhill, T. M., Chrysostomou, A., Hough, J. H., \& Yates, J. A.
  2000, \mnras, in press

\bibitem[Goebel \& Moseley\/(1985)]{goebel85}
  Goebel, J. H. \& Moseley, S. H.
  1985, \apj, 290, L35

\bibitem[von Helden et al.\/(2000)]{helden00}
  von Helden, G., Tielens, A. G. G. M., van Heijnsbergen, D.
  Duncanm M. A., Hony, S., Waters, L. B. F. M., \& Meijer, G.
  2000, Science, 288, 313  

\bibitem[Hill, Jones, \& d'Hendecourt\/(1998)]{hill98}
  Hill, H. G. M., Jones, A. P., \& d'Hendecourt, L. B.
  1998, \aap, 336, L41

\bibitem[Hinz et al.\/(1998)]{hinz98}
  Hinz, P. M., Angel, R. P., Hoffmann, W. F., McCarthy, D. W.,
  McGuire, P. C., Cheselka, M., Hora, J. H., \& Woolf, N. J.
  1998, Nature, 395, 251

\bibitem[Hoffmann et al.\/(1998)]{hoffmann98}
  Hoffmann, W. F., Hora, J. L., Fazio, G. G., 
  Deutsch, L. K.  \&  Dayal, A. 
  1998,  in Infrared Astronomical Instrumentation, 
  ed. A. M. Fowler, Proc. SPIE 3354, 647, 658

\bibitem[Hora et al.\/(1996)]{hora96}
  Hora, J. L., Deutsch, L. K., Hoffmann, W. F., \&
  Fazio, G. G.
  1996, \aj, 112, 2064
  
\bibitem[Hrivnak\/(1995)]{hrivnak95}
  Hrivnak, B. J.
  1995, \apj, 438, 341

\bibitem[Hrivnak \& Kwok\/(1991)]{hrivnak91}
  Hrivnak, B. J. \& Kwok, S.
  1991, \apj, 371, 631  

\bibitem[Hrivnak, Kwok, \& Geballe\/(1994)]{hkg94}
  Hrivnak, B. J., Kwok, S., \& Geballe, T. R.
  1994, \apj, 420, 783
  
\bibitem[Hrivnak et al.\/(1999)]{hrivnak99}
  Hrivnak, B. J., Langill, P. P., Su, K. Y. L.,
  \& Kwok, S
  1999, \apj, 513, 421

\bibitem[Iben\/(1995)]{iben95}
  Iben, I., Jr.
  1995, Phyics Reports, 250, 1

\bibitem[Jura\/(1986)]{jura86}
  Jura, M.
  1986, \apj, 303, 327

\bibitem[Jura\/(1994)]{jura94}
  Jura, M.
  1994, \apj, 434, 713

\bibitem[Jura, Chen, \& Werner\/(2000)]{jura01}
  Jura, M., Chen, C., \&  Werner, M. W.
  2000, \apj, 544, L141

\bibitem[Jura, Turner, \& Van Dyk\/(2000)]{jura00}
  Jura, M., Turner, J. L., \& Van Dyk, S.
  2000, \apj, 528, L105

\bibitem[Jura \& Werner\/(1999)]{jura99}
  Jura, M. \& Werner, M. W.
  1999, \apj, 525, L113

\bibitem[Justtanont et al.\/(1996)]{justtanont96}
  Justtanont, K., Barlow, M. J., Skinner, C. J., Roche, P. F., 
  Aitken, D. K., \& Smith, C. H.
  1996, \aap, 309, 612

\bibitem[Kim, Martin, \& Hendry\/(1994)]{kim94}
  Kim, S.-H., Martin, P. G., \& Hendry, P. D.
  1994, \apj, 422, 164

\bibitem[Knapp et al.\/(1994)]{knapp94}
  Knapp, G. R., Bowers, P. F., Young, K., \& Phillips, T. G.
  1994, , \apj, 429, L33

\bibitem[Kr\"uger \& Sedlmayr\/(1997)]{kruger97}
  Kr\"uger, D. \& Sedlmayr, E.
  1997, \aap, 321, 557
  
\bibitem[Kwok\/(1982)]{kwok82} 
  Kwok, S.
  1982, \apj, 258, 280
  
\bibitem[Kwok\/(1993)]{kwok93}
  Kwok ,S.
  1993, \araa, 31, 63

\bibitem[Kwok, Su, \& Hrivnak\/(1998)]{kwok98}
  Kwok, S., Su, K. Y. L., \& Hrivnak, B. J.
  1998, \apj, 501, L117

\bibitem[Kwok, Volk, \& Hrivnak\/(1989)]{kvh89}
  Kwok, S., Volk, K., \& Hrivnak, B. J.
  1989, \apj, 345, L51

\bibitem[Kwok, Volk, \& Hrivnak\/(1999)]{kvh99}
  Kwok, S., Volk, K., \& Hrivnak, B. J.
  1999, in IAU Symp. 191, Asymptotic Giant Branch Stars,
  ed. T. Le Bertre, A. Lebre, \& C. Waelkens 
  (San Fransisco: ASP)

\bibitem[Lindqvist et al.\/(1988)]{lind88}
   Lindqvist M., Nyman L.-A., Olofsson H., Winnberg A. 
   1988, 205, L15

\bibitem[Manchado et al.\/(1989)]{manchado89}
  Manchado, A., Pottasch, S. R., Garc\'{\i}a-Lario, P.,
  Esteban, C., \& Mampaso, A.
  1989, \aap, 214, 139

\bibitem[Meixner et al.\/(1997)]{meixner97}
  Meixner, M., Skinner, C. J., Graham, J. R., Keto, E., 
  Jernigan, J. G., \& Arens, J. F.
  1997, \apj, 482, 897

\bibitem[Meixner et al.\/(1999)]{meixner99}
  Meixner, M., Ueta, T., Dayal, A., Hora, J. H., Fazio, G., 
  Hrivnak, B. J., Skinner, C. J., Hoffman, W. F., \&
  Deutsch, L. K.
  1999, \apjs, 122, 221

\bibitem[Meixner, Ueta, \& Bobrowsky\/(2001)]{meixner00}
  Meixner, M., Ueta, \& Bobrowsky, M.
  2001, in preparation

\bibitem[Meixner, Young Owl, \& Leach (1999)]{nirim}
  Meixner, M., Young Owl, L., \& Leach, R.
  1999, \pasp, 111, 997

  \bibitem[Mastrodemos \& Morris\/(1999)]{mm99}
  Mastrodemos, N. \& Morris, M.
  1999, \apj, 523, 357

\bibitem[Morris \& Sahai\/(2000)]{morris00}
  Morris, M. \& Sahai, R.
  2000, in Asymmetrical Planetary Nebulae II: From Origins to
  Microstructures, ed. J. H. Kastner, N. Soker, \& S. A. Rappaport
  (San Fransisco: ASP), 143

\bibitem[Neckel \& Klare\/(1980)]{neckel80}
  Neckel, Th., \& Klare, G.
  1980, \aaps, 42, 251
  
\bibitem[Nuth et al.\/(1985)]{nuth85}
  Nuth, J. A., Moseley, S. H., Silverberg, R. F., 
  Goebel, J. H., \& Moore, W. J.
  1985, \apj, 290, L41

\bibitem[Omont et al.\/(1993)]{omont93}
  Omont, A., Loup C., Forveille T., te Lintel Hekkert P., 
  Habing H., Sivagnanam P. 
  1993, \aap, 267, 515

\bibitem[Omont et al.\/(1995)]{omont95}
  Omont, A., Moseley, S. H., Cox, P., Glaccum, W., Casey, S., 
  Forveille T., Chan, K.-W., Szczerba, R., Loewenstein, R. F., 
  Harvey, P. M., \& Kwok, S.
  1995, \apj, 454, 819

\bibitem[Pottasch \& Parthasarathy\/(1988)]{pottasch88}
  Pottasch, S. R., Parthasarathy, M.
  1988, \aap, 192, 182

\bibitem[Ressler et al.\/(1994)]{ressler94}
  Ressler, M. E., Werner, M. W., Van Cleve, J., \& 
  Choa, H. 
  1994, Exp. Astron., 3, 277 

\bibitem[Sahai et al.\/(1998)]{sahai98}
  Sahai, R., Hines, D. C., Kastner, J. H., Weintraub, D. A., 
  Trauger, J. T., Rieke, M. J., Thompson, R. I., 
  \& Schneider, G.
  1998, \apj, 492, L163

\bibitem[Skinner et al.\/(1994)]{skinner94}
  Skinner, C. J., Meixner, M., Hawkins, G. W., 
  Keto, E., Jernigan, J. G., \& Arens, J. F.
  1994, \apj, 423, L135

\bibitem[Skinner et al.\/(1997)]{skinner97}
  Skinner, C. J., Meixner, M., Barlow, M. J., Collison, A. J.,
  Justtanont, K., Blanco, P., Pina, R., Ball, J. R., Keto, E.,
  Arens, J. F., \& Jernigan, J. G.
  1997, \aap, 3328, 290

\bibitem[Soker\/(1997)]{soker97}
  Soker, N.
  1997, \apjs, 112, 487

\bibitem[Su et al.\/(1998)]{su98}
  Su, K. Y. L., Volk, K., Kwok, S., \& Hrivnak, B. J.
  1998, \apj, 508, 744

\bibitem[Szczerba et al.\/(1997)]{szczerba97}
  Szczerba, R., Omont, A., Volk, K., Cox, P., \& Kwok, S.
  1997, \aap, 317, 859

\bibitem[Ueta, Meixner, \& Bobrowsky\/(2000)]{ueta00}
  Ueta, T., Meixner, M., \& Bobrowsky, M.
  2000, \apj, 528, 861
    
\bibitem[Ueta et al.\/(2001)]{ueta01}
  Ueta, T., Meixner, M., Dayal, A., Deutsch, L. K., Fazio, G. G.,
  Hora, J. L., \& Hoffmann, W. F.
  2001, \apj, 548, 1020

\bibitem[van der Veen, Habing, \& Geballe\/(1989)]{vhg89}
  van der Veen, W. E. C. J., Habing, H. J.,  \& 
  Geballe, T. R.
  1989, \aap, 226, 108

\bibitem[Volk, Kwok, \& Hrivnak\/(1999)]{vkh99}
  Volk, K., Kwok, S., \& Hrivnak, B. J.
  1999, \apj, 516, L99

\bibitem[West et al.\/(1997)]{west97}
  West, S. C., Callahan, S., Chaffee, F. H., 
  Davison, W. B., Derigne, S. T., Fabricant, D. G., 
  Foltz, C. B., Hill, J. M., Nagel, R. H., Poyner, A. D., 
  \& Williams, J. T.
  1997, SPIE, 2871, 38

\bibitem[Woodsworth, Kwok, \& Chan\/(1990)]{woodsworth90}
  Woods
worth, A. W., Kwok, S., \& Chan, S. J.
  1990, \aap, 228, 503

\bibitem[Yuasa, Unno, \& Magono\/(1999)]{yuasa99}
  Yuasa, M., Unno, W., \& Magono, S.
  1999, \pasj, 51, 197

\bibitem[Za\v{c}s, Klochkova, \& Panchuk\/(1995)]{zacs95}
  Za\v{c}s, L., Klochkova, V. G., \& Panchuk, V. E.
  1995, \aap, 275, 764

\bibitem[Zubko et al.\/(1996)]{zubko96}
  Zubko, V. G., Mennella, V., Colangeli, L., \& Bussoletti, E.
  1993, \mnras, 282, 1321

\bibitem[Zuckerman \& Dyck\/(1986)]{zd86}
  Zuckerman, B. \& Dyck, H. M.
  1986, \apj, 311, 345

\end{thebibliography}
\end{document}